\documentclass[12pt]{article}

\makeatletter
\newfont{\footsc}{cmcsc10 at 8truept}
\newfont{\footbf}{cmbx10 at 8truept}
\newfont{\footrm}{cmr10 at 10truept}
 \makeatother 

\begin{document}


\title{Moments of characteristic polynomials enumerate two-rowed lexicographic arrays}%
\author{E. Strahov \\
\small Department of Mathematical Sciences,\\[-0.8ex]
\small Brunel University, Uxbridge,
 UB8 3PH, United Kingdom \\[-0.8ex]
\small \texttt{Eugene.Strahov@brunel.ac.uk}}

\date{21.11.2001}%
\maketitle

\begin{abstract}
 A combinatorial interpretation is provided for the moments of characteristic
polynomials of random unitary matrices. This leads to a rather
unexpected consequence of the Keating and Snaith conjecture: the
moments of $\mid\xi(1/2+it)\mid$ turn out to be connected with
some increasing subsequences problems (such as the last passage
percolation problem).
\end{abstract}

\section{Introduction}
Keating and Snaith \cite{keating1} have proposed to model the
limiting distribution of the Riemann zeros using the
characteristic polynomials of unitary random matrices $U$
\begin{equation}
Z(U,\theta)=\textmd{det}(I-e^{-i\theta}U)
\end{equation}
In particular they deal with a conjecture for moments of
$|\zeta(1/2+it)|$ which states that the limit
\begin{equation}\label{1.2}
I_m=\lim\limits_{T\rightarrow\infty} \frac{1}{(\log
T)^{m^2}}\int\limits_0^T|\zeta(1/2+it)|^{2m}dt
\end{equation}
exists and is equal to a product of two factors, $f(m)$ and
$a(m)$, i. e.
\begin{equation}\label{1.3}
I_m=a(m)f(m)
\end{equation}
The first factor $a(m)$ is the zeta-function specific part,
\begin{equation}\label{f}
a(m)=\prod\limits_{p}
(1-\frac{1}{p})^{m^2}\sum_{k=0}^{+\infty}\left(\frac{\Gamma(k+m)}{k!\mbox{
} \Gamma(m)}\right)^2 p^{-k}
\end{equation}
where the product is taken over prime numbers $p$. As for the
second factor $f(m)$, Keating and Snaith \cite{keating1} have
hypothesized that it is the random matrix (universal) part and may
be represented as
\begin{equation}\label{fm}
f(m)=\lim\limits_{N\rightarrow\infty}N^{-m^2}
\langle\left|Z(U,\theta)\right|^{2m}\rangle_{U(N)}
\end{equation}
Here the average is over Circular Unitary Ensemble (CUE) of random
unitary matrices $N\times N$.

In this letter I provide a combinatorial interpretation for the
moments of characteristic polynomials,
$\langle\left|Z(U,\theta)\right|^{2m}\rangle_{U(N)}$.  I then
relate these moments with two-rowed lexicographic arrays which are
generalizations of permutations and words in combinatorics (for
basic information about permutations, words and lexicographic
arrays see, for example, a book by Fulton\cite{fulton}). The
combinatorial interpretation of moments of the characteristic
polynomials enables  an explicit formula to be obtained for the
total number of two-rowed lexicographic arrays constructed from
the letters of an alphabet of $m$ symbols with the increasing
subsequences of the length at most $N$.

This combinatorial interpretation is, in fact, a natural
consequence of a profound relation between Random Matrix Theory
and Robinson-Schensted-Knuth type problems  discovered recently by
Gessel \cite{gessel}, Baik, Deift and Johansson \cite{baik1} and
Rains \cite{rains} (see also Aldous and Diaconis \cite{aldous} and
the references therein ). In particular, it follows that certain
expectation values over CUE appear in the theory of last passage
percolation. The moments of characteristic polynomials ,
$\langle\left|Z(U,\theta)\right|^{2m}\rangle_{U(N)}$, may also be
considered in this context. The reasoning along those lines
enables the random matrix part $f(m)$ defined by equations
(\ref{1.2})-(\ref{f}) to be related with the
weakly-right/weakly-up lattice model of the last passage
percolation.
\section{Increasing subsequences of lexicographic arrays and
$\langle\left|Z(U,\theta)\right|^{2m}\rangle_{U(N)}$ } Let us
recall a generalization of the Cauchy identity (see Gessel
\cite{gessel}, Baik and Rains \cite{baik2}, Tracy and Widom
\cite{tracy} ):
\begin{eqnarray}\label{schurexpansion}
\sum\limits_{N\geq\lambda_1\ldots\geq\lambda_m\geq
0}s_{\lambda}\left(\xi_1,\xi_2,\ldots ,\xi_m\right)
s_{\lambda}\left(\eta_1,\eta_2,\ldots ,\eta_m\right)= \nonumber \\
\left\langle\det\left(\prod\limits_{i,j=1}^{m}\left(I+
\xi_iU\right)\left(I+
\eta_jU^{\dag}\right)\right)\right\rangle_{U(N)}
\end{eqnarray}
Here $s_{\lambda}\left(\xi_1,\xi_2,\ldots ,\xi_m\right)$ denotes
the Schur polynomial of $m$ variables. Take
$\xi_1=\xi_2=\ldots=\xi_m=e^{-i\theta}$ and
$\eta_1=\eta_2=\ldots=\eta_m=e^{i\theta}$ in  equation
(\ref{schurexpansion}). We will use the formula for the Schur
polynomials of identical variables,
\begin{equation}\label{schuridentical}
s_{\lambda}(x,x,\ldots,x)=x^{\lambda_1+\lambda_2+\ldots+\lambda_m}d_{\lambda}(m)
\end{equation}
where the coefficient $d_{\lambda}(m)$ is calculated by the
formula
\begin{equation}\label{dlambda}
d_{\lambda}(m)=\frac{\prod\limits_{1\leq i<j\leq
m}\left(\lambda_i-i-\lambda_j+j\right)}{0!\mbox{ }1!\mbox{
}2!\ldots(m-1)!}
\end{equation}
From equations (\ref{schurexpansion})-(\ref{dlambda}) we conclude
that the moments of characteristic polynomials,\linebreak
$\langle\left|Z(U,\theta)\right|^{2m}\rangle_{U(N)}$, may be
represented as sums over partitions. Importantly, these sums
should be taken only over partitions of the length less than $m$
and with the first row less than $N$. An explicit expression for
the $m^{\mbox{th}}$ moment of the characteristic polynomial is
\begin{equation}\label{momentsexpansion}
\langle\left|Z(U,\theta)\right|^{2m}\rangle_{U(N)}=
\sum\limits_{K=0}^{+\infty}\mbox{ }\sum\limits_{\lambda\vdash
K,\mbox{ }\lambda_1\leq N} d_{\lambda}^2(m)
\end{equation}
where $\lambda\vdash K$ means that the set $(\lambda_1,
\lambda_2,\ldots ,\lambda_m)$ is a partition of $K$, i.e.
$\lambda_1+\lambda_2+\ldots+\lambda_m=K$. Representation
(\ref{momentsexpansion}) enables  a combinatorial interpretation
to be provide for
$\langle\left|Z(U,\theta)\right|^{2m}\rangle_{U(N)}$.

Consider semi-standard Young tableaux constructed from $K$ boxes
with at most $m$ and $N$ boxes in first columns and first rows
respectively. The total number of pairs of such tableaux is then
given by the sum $\sum\limits_{\lambda\vdash K,\mbox{
}\lambda_1\leq N} d_{\lambda}^2(m)$. (It is a well-known fact (
see Fulton \cite{fulton}) that the coefficient $d_{\lambda}(m)$
defined by equation (\ref{dlambda}) gives the number of
semi-standard Young tableaux of the shape $\lambda$ whose entries
should be taken from the set $[1, 2,\ldots , m]$). It is this sum
which appears in expression (\ref{momentsexpansion}) for the
moments of the characteristic polynomials. In what follows we will
apply the Robinson-Schensted-Knuth correspondence (see Fulton
\cite{fulton}) between two-rowed lexicographic arrays and pairs of
semi-standard Young tableaux.

By definition, a two-rowed array of the size $K$ is an object of
the form
\begin{equation}\label{arrayform}
\emph{A}_{2,K}^{m}= \left(\begin{array}{cccc}
u_1 & u_2 & \ldots & u_K \\
v_1 & v_2 & \ldots & v_K
\end{array} \right)
\end{equation}
where the letters $u_1, u_2,\ldots ,u_K$ and $v_1, v_2,\ldots
,v_K$ belong to an alphabet (any ordered set) $\aleph_m$ of $m$
different letters. The following two conditions ensure that the
array $\emph{A}_{2,K}^m$ corresponds to a pair of semi-standard
Young tableaux (each of $K$ boxes with entries taken from
$1,2,\dots, m$):
\begin{equation}\label{lcondition1}
u_1\leq u_2\leq\ldots\leq u_K
\end{equation}
\begin{equation}\label{lcondition2}
v_1\leq v_2\leq\ldots\leq v_K
\end{equation}
The arrays $\emph{A}_{2,K}^m$ for which  conditions
(\ref{lcondition1}) and (\ref{lcondition2}) are satisfied are
called lexicographic arrays.

Let us define a (weakly) increasing subsequence of the
lexicographic array $\emph{A}_{2,K}^m$ as follows:
\begin{enumerate}
  \item The element of
the array $\emph{A}_{2,K}^m$ with the number $i_1$ is less than
that with the number $i_2$, ${u_{i_1}\choose v_{i_1}}\leq
{u_{i_2}\choose v_{i_2}}$, if $v_{i_1}\leq v_{i_2}$ and
$u_{i_1}\leq u_{i_2}$.
  \item  A subsequence of  $s$ elements of the array $\emph{A}^{m}_{2,K}$
such that
\begin{eqnarray}
{u_{i_1}\choose v_{i_1}}\leq{u_{i_2}\choose v_{i_2}}
\leq\ldots\leq{u_{i_s}\choose v_{i_s}}, \quad i_1<i_2<\ldots< i_k
\end{eqnarray}
will be called a weakly increasing subsequence of the length $s$.
\end{enumerate}
Let $l(\emph{A}_{2,K}^m)$ be the length of the maximal (weakly)
increasing subsequence of the two-rowed lexicographic array
$\emph{A}_{2,K}^m$. Denote by $\emph{R}^K_{m, N}$ the number of
two-rowed lexicographic arrays of the size $K$ constructed from
the alphabet of $m$ letters $\aleph_m$ and including weakly
increasing subsequences of the length at most $N$, i.e.
\begin{equation}\label{rmnk}
\emph{R}^K_{m,N}= \sharp \mbox{ of } \emph{A}_{2,K}^m \;\;
\mbox{with}\quad l(\emph{A}_{2,K}^m)\leq N
\end{equation}
The number $\emph{R}_{m, N}$ defined by the sum over size $K$,
\begin{equation}\label{rmn0}
\emph{R}_{m,N} = \sum\limits^{+\infty}_{K=0} \emph{R}^K_{m,N}
\end{equation}
gives the number of arrays of an arbitrary size that include
weakly increasing subsequences of $N$ elements or less. By the
Robinson-Schensted-Knuth correspondence, the number of ordered
pairs of semi-standard Young tableaux with $m$ and $N$ boxes in
the first column and the first row respectively is equal to the
number $\emph{R}_{m, N}$ of lexicographic arrays.

This observation enables  $\emph{R}_{m, N}$ to be expressed as an
average over CUE
\begin{equation}\label{rmn}
\emph{R}_{m, N}=\sum\limits_{K=0}^{+\infty}\mbox{
}\sum\limits_{\lambda\vdash K,\mbox{ }\lambda_1\leq N}
d_{\lambda}^2(m)=\langle\left|Z(U,\theta)\right|^{2m}\rangle_{U(N)}
\end{equation}
Here we have used the representation of the moments
$\langle\left|Z(U,\theta)\right|^{2m}\rangle_{U(N)}$ as the sums
over partitions, see equation (\ref{momentsexpansion}). We thus
conclude that the moments of characteristic polynomials,
$\langle\left|Z(U,\theta)\right|^{2m}\rangle_{U(N)}$, have a clear
combinatorial interpretation. They are equal to the number of
two-rowed lexicographic arrays constructed from an alphabet of $m$
symbols with the weakly increasing subsequences of the $N$
elements at most.

It is now possible to obtain an explicit formula for the number
$\emph{R}_{m, N}$ of lexicographic arrays. Once we apply the
result of Keating and Snaith\cite{keating1}:
\begin{equation}\label{keating}
\langle\left|Z(U,\theta)\right|^{2m}\rangle_{U(N)} =
\prod\limits_{j=1}^{N}\frac{\Gamma(j)\Gamma(j+2m)}{\left[\Gamma(j+m)\right]^2}
\end{equation}
equality (\ref{rmn}) between the number of arrays $\emph{R}_{m,
N}$ and the moments of characteristic polynomials gives
\begin{equation}\label{rmn1}
\emph{R}_{m,
N}=\prod\limits_{j=1}^{N}\frac{\Gamma(j)\Gamma(j+2m)}{\left[\Gamma(j+m)\right]^2}
\end{equation}

In order to illustrate how formula (\ref{rmn1}) works consider the
following example. Take an alphabet consisting of two letters $a$
and $b$, i.e. $\aleph_m=(a,b),\quad m=2$. It may give rise to both
lexicographic and non-lexicographic arrays. A typical
lexicographic array compounded from the letters of the alphabet
$(a,b)$ is
\begin{displaymath}
\left(\begin{array}{cccc}
a & a & b & b \\
a & b & b & b
\end{array}\right)
\end{displaymath}
We observe that conditions (\ref{lcondition1}),
(\ref{lcondition2}) are satisfied by this particular array. An
example of a two-rowed non-lexicographic array is given below:
\begin{displaymath}
\left(\begin{array}{cccc}
a & a & b & b \\
a & b & a & b
\end{array} \right)
\end{displaymath}
In this array the second letter of the word $a b a b$, $b$, is
larger than the third letter of this word $a$. Thus, condition
(\ref{lcondition2}) is not satisfied resulting in a
non-lexicographic array.

Let us list all possible two-rowed lexicographic arrays from the
alphabet $(a,b)$ with the weakly increasing subsequences of the
length at most $N=2$:
\begin{eqnarray}
{\o\choose\o}\;\quad {a \choose a}\;\quad {a \choose b}\;\quad {b
\choose b}\;\quad {b \choose a}\;\quad {a  b \choose a  a}\;\quad
{a a \choose a  b}\;\quad {a  b \choose b  a}\;\quad {a  a \choose
b b}\;\quad {a  a \choose a  a} \hspace{0.5cm}\nonumber \\
{b  b \choose a b} \quad {b b \choose b  b}\quad {a b \choose b
b}\quad {b  b \choose a a}\quad  {a  b \choose a b}\quad {a  b b
\choose b a b}\quad {a a  b \choose b  b a}\quad {a  b  b \choose
b  a a}\quad {a a b \choose a  b a}\quad {a  a  b  b \choose b b a
a}\nonumber 
\end{eqnarray}
There are 20 arrays with the required properties. It can be easily
verified that   formula (\ref{rmn1}) obtained above gives
precisely this number, i.e.
\begin{equation}
\emph{R}_{m, N}=\prod_{j=1}^{N}
\frac{\Gamma(j)\Gamma(j+2m)}{\lbrack\Gamma(j+m)\rbrack^2}
=20\qquad (m=2, N=2)
\end{equation}
\section{Moments of characteristic polynomials in the last passage
percolation theory} Let us now turn to an interesting
interpretation of lexicographic arrays in the framework of the
last passage percolation theory. (See, for example, Baik
\cite{baik3} where different last passage percolation models are
described and the relations with different expectation values over
CUE are explained). It is a consequence of the relation between
the moments of characteristic polynomials and lexicographic arrays
that $\langle\left|Z(U,\theta)\right|^{2m}\rangle_{U(N)}$ appear
also in the last passage percolation problems.

Consider the lexicographic array $\emph{A}_{2,K}^m$ of the form
(\ref{arrayform}) where the letters $u_i, v_i; i=1,2,\ldots ,K$
take values in the set of integers $\aleph_m={1,2,\ldots ,m}$.
This lexicographic array corresponds to a matrix with non-negative
integer entries. The $(i,j)$ entry of this matrix is the number of
times the vector ${i\choose j}$ occurs in the array
$\emph{A}_{2,K}^m$. For example, the array
\begin{displaymath}
\left(\begin{array}{cccccccccc}
1 & 1 & 1 & 2 & 2 & 2 & 3 & 3 & 3 & 3 \\
1 & 2 & 2 & 1 & 1 & 1 & 1 & 2 & 3 & 3 \\
\end{array} \right)
\end{displaymath}
corresponds to the matrix
\begin{displaymath}
\left(
\begin{array}{ccc}
  1 & 2 & 0 \\
  3 & 0 & 0 \\
  1 & 1 & 2
\end{array}
\right)
\end{displaymath}
Let $X(i,j)$, $i,j\in\left[1,2,\ldots ,m\right]$ be a planar array
of non-negative integers. Consider weakly-up/weakly-right paths
$\pi'^s$, ${(i_k,j_k)}^l_{k=1}$, such that $i_1\leq
i_2\leq\ldots\leq i_l$ and $j_1\leq j_2\leq\ldots\leq j_l$ (This
model has  discussed at length in Baik \cite{baik3}\cite{baik4}).
Let $(1,1)\nearrow(m,m)$ denote the set of up/right paths from
$(1,1)$ to $(m,m)$ in the planar array $X(i,j)$. The entries of
the matrix $X(i,j)$, $x_{ij}$, may be considered as the passage
times at knots $(i,j)$. Each planar array of non-negative integers
$X(i,j)$ can be assigned the last passage percolation time
$T(m,m)$ for  travel from $(1,1)$ to $(m,m)$. $T(m,m)$ is defined
explicitly by the following expression:
\begin{equation}\label{percolationtime}
T(m,m)=\max\limits_{\pi\in(1,1)\nearrow(m,m)}\sum\limits_{i,j\in\pi}x_{ij}
\end{equation}
So far we have used the Robinson-Schensted-Knuth correspondence
between pairs of semi-standard Young diagrams and lexicographic
arrays. Turning to the last passage percolation problems
Robinson-Schensted-Knuth correspondence is a correspondence
between planar arrays $X(i,j)$ of non-negative integers and pairs
of semi-standard Young tableaux. In this case the last passage
percolation time $T(m,m)$ should be considered instead of the
length of the longest increasing subsequence. The number of arrays
$\emph{R}_{m,N}$ defined in the previous section by equations
(\ref{rmnk}) and (\ref{rmn0}) is replaced by the number of all
possible $X(i,j)$ of size $m\times m$ with the last passage
percolation time $T(m,m)$ less than $N$. Applying the results of
the previous section we find that
\begin{equation}\label{planarm}
\sharp\;\mbox{of}\; X(i,j)\;\mbox{with}\; T(m,m)\leq
N=\langle\left|Z(U,\theta)\right|^{2m}\rangle_{U(N)}
\end{equation}
With the help of the Keating and Snaith formula (equation
(\ref{keating}))  an explicit expression is obtained for the
number of planar arrays $X(i,j)$ of the size $m\times m$ with the
last passage percolation time bounded by $N$:
\begin{equation}\label{planarg}
\sharp\;\mbox{of}\; X(i,j)\;\mbox{with}\; T(m,m)\leq
N=\prod\limits_{j=1}^{N}\frac{\Gamma(j)\Gamma(j+2m)}{\left[\Gamma(j+m)\right]^2}
\end{equation}
\section{Combinatorial consequences of the Keating and Snaith conjecture}
The above results lead to  an interesting interpretation of the
Keating and Snaith conjecture. Once $m$ is a non-negative integer,
the random matrix (universal) factor $f(m)$ relates the moments
$I_m$ of zeta-function with  increasing subsequences. On
assumption that the Keating and Snaith conjecture is valid, that
is equation (\ref{fm}) holds, it follows from the results of the
previous sections that
\begin{equation}
\frac{I_m}{a(m)}=\lim\limits_{N\rightarrow\infty}\frac{\sum\limits_{K=0}^{\infty}
\sharp\;\mbox{of}\;\emph{A}^m_{2,K}\;\mbox{with}\;l(A_{2,K}^m)\leq
N}{N^{m^2}}
\end{equation}
In other words, the ratio $\frac{I_m}{a(m)}$ defined by equations
(\ref{1.2})-(\ref{f}) is asymptotically equal to the number of
lexicographic arrays constructed from an alphabet of $m$ letters
with increasing subsequences of at most $N$ elements divided by
$N^{m^2}$.

As a particular case consider  the  weakly-up/weakly-right last
passage percolation model. Here the ratio $\frac{I_m}{a(m)}$ is
equal to the whole number of $m\times m$ planar arrays $X(i,j)$ of
non-negative integers with last percolation time at most $N$
divided by $N^{m^2}$:
\begin{equation}
\frac{I_m}{a(m)}=\lim\limits_{N\rightarrow\infty}
\frac{\sharp\;\mbox{of}\;X(i,j)\;\mbox{with}\;T(m,m)\leq
N}{N^{m^2}}
\end{equation}
Apart from the combinatorial interpretation of the Keating and
Snaith conjecture, the  above considerations raise a question of
how the moments of the Riemann zeta function are related to the
representation theory of finite groups.
\subsection*{Acknowledgements.}
For valuable comments and discussions that have contributed to
this work I am grateful to Yan Fyodorov. This research was
supported by EPSRC grant GR/R13838/01 "Random Matrices close to
Unitary or Hermitian."

\end{document}